\input{aipcheck}

\documentclass[
    ,final            
  ]
  {aipproc}

\layoutstyle{6x9}

\begin{document}

\title{Astrophysical Implication of Low  E($2^+_1$) in Neutron-rich Sn Isotopes}


\author{S. Sarkar}{
  address={Bengal Engineering and Science University, Shibpur, Howrah - 711103, INDIA}
}

\author{M. Saha Sarkar}{
  address={Saha Institute of Nuclear Physics, Kolkata - 700064, INDIA}
}

\begin{abstract}
The observation and prediction of unusually depressed first excited 
2$^+_1$ states in even-A neutron - rich isotopes of semi-magic Sn above 
$^{132}$Sn provide motivations for reviewing the problems related to the 
nuclear astrophysics in general. In the present work, the $\beta$-decay rates 
of the exotic even Sn isotopes ($^{134,136}$Sn) above the $^{132}$Sn 
core have been calculated as a function of temperature (T). In order to get 
the necessary ft values, B(GT) values corresponding to allowed Gamow Teller 
(GT$-$) $\beta$-decay have been  theoretically calculated using shell model. 
The total decay rate shows decrease with increasing temperature as the 
ground state population is depleted and population of excited states with 
slower decay rates increases. The abundance at each Z value is inversely 
proportional to the decay constant of the waiting point nucleus for that 
particular Z. So the increase in half-life of isotopes of Sn, like $^{136}$Sn, 
might have substantial impact on the r-process nucleosynthesis.

\end{abstract}

\maketitle


\section{INTRODUCTION}

Nuclei with $50\leq$Z$\leq  56$ and  82 $\leq$ N $\leq$ 88 in the  
$\pi(gdsh)  \oplus \nu(hfpi)$  valence  space  above  the $^{132}Sn$ core 
lie on or close to the path of astrophysical r-process flow. Their structure, 
particularly the binding energy (BE), low-lying excited states and 
beta decay rates at finite temperatures are important ingredients for 
nucleosynthesis calculations. Sn isotopes are of particular importance. 
Even Sn isotopes, say $^{136}Sn$, is known \cite{woh:1} to be the classical "waiting point" 
nucleus in A=130 solar system abundance peak under typical r-process condition.
Spectroscopic information, such as BE and low lying spectrum, is  known 
experimentally only for $^{134}Sn$ \cite{nndc}. Half-lives of $^{135-137}Sn$ have 
been measured through $\beta^-n$  decay process \cite{sher}. 
No other information exists. Lifetimes of these nuclei are very small and 
production rate very low presenting challenges to spectroscopic studies. 
Reliable theoretical results are therefore necessary and useful.\\

In general, the $\beta^-$-decay rates ($\lambda$)  relevant in the astrophysical scenario
depend on the density ($\rho$), electron fraction  (Y$_e$)  and the prevalent 
temperature ($T$) in the environment \cite{kar}. In the present work, the decay rates of the exotic Sn isotopes
($^{134,136}Sn$) above the $^{132}Sn$ core have been calculated as a function 
of temperature (T) only. The necessary $ft$ 
values have been obtained from the B(GT)s, corresponding to allowed Gamow 
Teller (GT$_-$) $\beta^-$ -decay   using untruncated shell model calculations.

\begin{figure}
\includegraphics[height=.2\textheight, viewport=00 20 410 550]{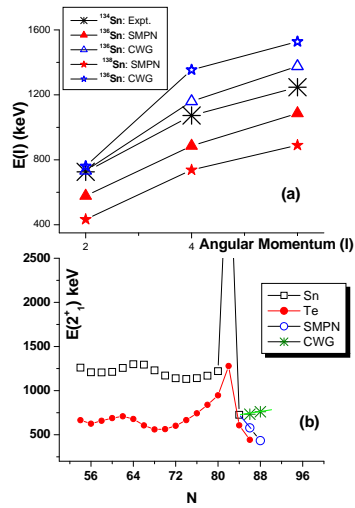}
  \caption{(a) Predictions for excitation energies in $^{136,138}Sn$ with CWG and SMPN interactions. The experimental spectrum for $^{134}Sn$ is also shown in the figure.(b)  Casten-Sherrill systematics.}
\end{figure}

\section{CALCULATIONS}

Untruncated shell model calculations in the valence space consisting of $\pi(1g_{7/2}, ~ 2d_{5/2}$, $2d_{3/2},   ~
3s_{1/2},  1h_{11/2})$  and  $\nu(1h_{9/2}, ~ 2f_{7/2},  ~2f_{5/2}, ~3p_{3/2}, 3p_{1/2}, 1i_{13/2})$ orbitals 
with the empirical SMPN \cite{sar}  and effective CWG \cite{bro} (1+2) - body Hamiltonians
show very good agreement with the available experimental data. But it is remarkable
that for $^{136,138}Sn$, where experimental level schemes are not known, the theoretical predictions
differ dramatically \cite{prc}. For $^{134-138}Sn$ (Fig. 1a), CWG interaction predicts nearly
constant energies of 2$^+_1$ states, normally expected for semi-magic nuclei. But SMPN 
predicts a remarkable new feature: decreasing E(2$^+_1$ ) energies with increasing neutron number.
Casten and Sherrill \cite{cast} have pointed out that, although [E(2$^+_1$ )Sn - E(2$^+_1$ )Te] $\simeq$ 400
keV (Fig.1b) for a given neutron number over most of the N = 50 - 82 shell, the difference
[E(2$^+_1$ )Sn - E(2$^+_1$)Te] is only 119 keV for N = 84. It is indeed remarkable that the difference
[E(2$^+_1$)($^{136}Sn$) - E(2$^+_1$)($^{138}Te$)] for N = 86 is 108 keV with SMPN. So it is consistent with
the trend discussed by Casten and Sherrill. For CWG, this difference is 733 - 356 = 377 keV for N = 86, which deviates from
the trend.

This observation of depressed first excited states in even-A Sn isotopes provides very 
useful ingredient for reviewing the problems related to the nuclear astrophysics in general. 
In the present work  an estimation of the effect of the depressed energies on  the  $\beta$- decay rates  of the exotic even
Sn isotopes ($^{134,136}$Sn) above the $^{132}Sn$ core have been calculated as a function of temperature (T).  

In astrophysical environments, thermally populated excited states in the mother nucleus (namely, Sn isotope in the present situation) 
can have an equilibrium population. The excited levels  undergo $\beta$-decay transitions to the ground state or to excited states 
in the daughter nucleus (corresponding  Sb isotope). These additional $\beta$-decay transitions may alter the astrophysical 
half-lives of the Sn isotopes  compared to the laboratory values. The change in the half-life can in turn 
have an impact on the r process nucleosynthesis and generation of more exotic neutron rich species. 

In order to get the necessary $ft$ values corresponding to the decay of the thermally populated excited states of the mother 
to the excited states of the daughter nucleus only allowed Gamow -Teller (GT-) transitions have been considered. B(GT) values 
have been calculated using OXBASH code \cite{oxb} with both SMPN and CWG Hamiltonians. 
It is evident that the thermal population of excited nuclear levels becomes more important with increasing temperature 
and lower excitation energy. The beta decay rate for a nucleus in astrophysical environment at a temperaure T is given by \cite{kar}

\begin{equation}
\lambda= \sum_i (2I_i+1) e^{-E_i/kT} \sum_{j} \lambda_{ij}/G
\end{equation}

where $E_i$ is the energy of the state of the mother nucleus and G (=$\sum_{i} (2I_i+1)e^{-E_i/kT}$) is the partition function of
the mother nucleus. Index $j$ sums over states of the daughter nucleus to which transitions are allowed.
The rate from the parent nuclear state $i$ to the daughter nuclear state $j$ is given by

\begin{equation}
\lambda_{ij} = {ln 2 \over (ft)_{ij}}f_{ij}
\end{equation}

$f_{ij}$ is the phase-space factor \cite{gov} for beta decay. The  $(ft)_{ij}$ value of an allowed beta decay 
is given by \cite{kar}

\begin{equation}
(ft)_{ij} = {6250 s \over B(F)_{ij}+ (g_A/g_V)^2B(GT)_{ij}}
\end{equation}

where $g_A$,  $g_V$ are vector and axial-vector coupling constants. $B(F)_{ij}$ and $B(GT)_{ij}$ denote the Fermi (F) and 
Gamow-Teller (GT) transition probabilities from ith mother state to jth daughter state. 
 
\section{RESULTS AND DISCUSSIONS}

For both the isotopes of Sn, ground state to ground state beta decays are forbidden transitions. But they are quite fast.
The relevant information for these isotopes incorporated for modifications in the decay rates that result from inclusion of excited states due to thermal excitations are shown in Table 1. The selection rules for GT transtions only allow transitions from single particle $\nu 1h_{9/2}$ orbital to 
$\pi 1h_{11/2}$ orbital in this model space. But the wavefunction compositions of the relevant  low lying states in these isotopes of Sn and Sb have 
very small contribution from the shell model configurations involving these orbitals. So the calculated allowed GT strengths are generally very small.
\begin{figure}
\includegraphics[height=.2\textheight, viewport=00 20 410 550]{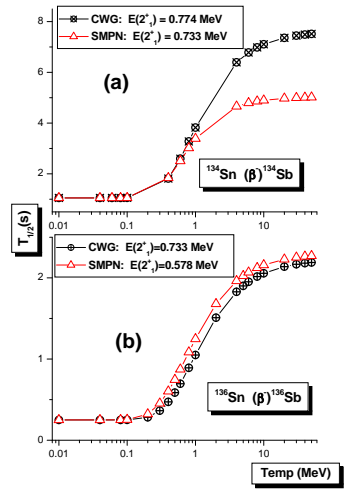}
  \caption{Variation of decay half life with temperature for (a)$^{134}Sn$ and (b)$^{136}Sn$.}
\end{figure}

For SMPN interaction, as discussed in Ref. \cite{sar}, the excitation energies of lowest yrast excited states of $^{134}Sn,Sb$ have been used for adjusting the two body neutron-neutron and neutron-proton matrix elements. So the energies of the $2^+_1$ states for $^{134}Sn$ are almost similar with the two interactions. But wavefunctions and energies predicted by CWG and SMPN are different for some of the states in $^{134}Sb$ resulting in different values of calculated B(GT)s. 

\begin{table}
\begin{tabular}{lccccc}
\hline
  \tablehead{1}{c}{b}{Mother}
  & \tablehead{1}{c}{b}{Half life \\Expt. (s)}
  & \tablehead{1}{c}{b}{Q\\value (MeV)}
  & \tablehead{1}{c}{b}{Mother ($Sn$)\\states}
  & \tablehead{1}{c}{b}{Daughter ($Sb$)\\states}   \\
\hline
$^{134}Sn$ &1.050& 7.37&2$^+_1$,2$^+_2$&3$^+_1$, 1$^+_1$, 2$^+_1$, 3$^+_2$\\
$^{136}Sn$ &0.250& 8.37&2$^+_1$,2$^+_2$&3$^+_1$, 1$^+_1$, 2$^+_1$, 3$^+_2$\\
\hline
\end{tabular}
\caption{Relevant information \cite{nndc} for calculations}
\label{tab:a}
\end{table}

With both  the interactions the effective half-life increases with increasing temperature for  $^{134,136}Sn$ (Fig.2).
Since the decay rate from the ground state is quite fast and  those from the first and second excited $2^+$ states are two orders of magnitude slower (for $kT$= 1 MeV), so the total decay rate decreases with increasing temperature as the ground state population is depleted and population of excited states with slower decay rates increases. 
For each case the extent and rate of increase depend on the details of the wavefunctions. At around 0.1 MeV temperature 
($\simeq 10^9$K), deviation from  the laboratory value of half life is inititated. In $^{134}Sn$ (Fig.2a) with both interactions, the
deviation starts at same temperature and the saturation is reached above 10 MeV. The saturated values of half lives with two interactions are different, $\simeq$ 5 s for SMPN and 
$\simeq$ 7.5 s for CWG. For $^{136}Sn$ (Fig.2b),  the relatively lower
E($2^+_1$) value (578 keV) with SMPN  compared to 733 keV with CWG is manifested by the faster rate of  change of T$_{1/2}$
with SMPN.  But both saturates
at $\simeq$ 2.2 s.

The abundance at each Z value is inversely proportional to the decay constant of the waiting point nucleus for that particular Z. So the increase in half-life of isotopes of Sn, like, $^{136}Sn$ will definitely have  substantial impact on the r-process nucleosynthesis.\\

\end{document}